\def\percc {$\hbox{{\rm cm}}^{-3}$}    
\def\kks {$\hbox{{\rm K km s}}^{-1}$}    
\def\kms {$\hbox{{\rm km s}}^{-1}$}    
\def\HH {\hbox{${\rm H}_2$}}  
\def\MOLN {\hbox{${\rm N}_2$}}  
\def\MOLH {\hbox{${\rm H}_2$}}  
\def\AMM {\hbox{${\rm NH}_{3}$}} 
\def\HCOP {\hbox{${\rm HCO}^+$}}      
\def\DCOP {\hbox{${\rm DCO}^+$}}    
\def\HTHP {\hbox{${\rm H}_{3}^{+}$}}   
\def\HTDP {\hbox{${\rm H}_{2}{\rm D}^{+}$}}   
\def\CEIO {\hbox{${\rm C}^{18}{\rm O}$}}   
\def\CSEO {\hbox{${\rm C}^{17}{\rm O}$}}   
\def\NTHP {\hbox{${\rm N}_2{\rm H}^+$}} 
\def\NTDP {\hbox{${\rm N}_2{\rm D}^+$}} 
\def\sec  {\hbox{$^{\prime \prime}$}}
\def\deg  {\hbox{$^{\rm o}$}}
\def\solmass {$\hbox{M}_\odot$}

\def\la{\mathrel{\mathchoice {\vcenter{\offinterlineskip\halign{\hfil
$\displaystyle##$\hfil\cr<\cr\sim\cr}}}
{\vcenter{\offinterlineskip\halign{\hfil$\textstyle##$\hfil\cr
<\cr\sim\cr}}}
{\vcenter{\offinterlineskip\halign{\hfil$\scriptstyle##$\hfil\cr
<\cr\sim\cr}}}
{\vcenter{\offinterlineskip\halign{\hfil$\scriptscriptstyle##$\hfil\cr
<\cr\sim\cr}}}}}
 
\def\ga{\mathrel{\mathchoice {\vcenter{\offinterlineskip\halign{\hfil
$\displaystyle##$\hfil\cr>\cr\sim\cr}}}
{\vcenter{\offinterlineskip\halign{\hfil$\textstyle##$\hfil\cr
>\cr\sim\cr}}}
{\vcenter{\offinterlineskip\halign{\hfil$\scriptstyle##$\hfil\cr
>\cr\sim\cr}}}
{\vcenter{\offinterlineskip\halign{\hfil$\scriptscriptstyle##$\hfil\cr
>\cr\sim\cr}}}}}

\documentclass{aa} 
\usepackage{graphics}

\begin{document}

\title{Observations of L1521F: a highly evolved starless core}

  \author{A. Crapsi\inst{1,2}
   \and P. Caselli\inst{3} 
   \and C.M. Walmsley\inst{3}
   \and M. Tafalla\inst{4}
   \and C.W. Lee\inst{5}
   \and T.L. Bourke\inst{6}
   \and P.C. Myers\inst{2}}

\offprints{}

\institute{Universit\'a degli Studi di Firenze Dipartimento di Astronomia e 
Scienza dello Spazio, , Largo E. Fermi 5, I-50125 Firenze, Italy
\and Harvard--Smithsonian Center for Astrophysics, 60 Garden Street, Cambridge, 
MA 02138, USA
\and INAF--Osservatorio Astrofisico di Arcetri, Largo E. Fermi 5, I-50125 Firenze, Italy
\and Observatorio Astron\'omico Nacional (IGN), Alfonso XII, 3, E-28014 Madrid, Spain
\and Taeduk Radio Astronomy Observatory, Korea Astronomy Observatory, 61-1 Hwaam-dong, 
Yusung-gu, Daejon 305-348, Korea
\and Harvard--Smithsonian Center for Astrophysics, Submillimeter Array Project, 645 N.
A'ohoku Place, Hilo, HI 96720, USA 
}

\date{Received ;Accepted}
\titlerunning{Observations of L1521F}

\abstract { We observed the pre--stellar core L1521F in dust
emission at 1.2mm and in two transitions each of \NTHP, \NTDP, \CEIO \
and \CSEO \ in order to increase the sample of well studied
centrally concentrated and chemically
evolved starless cores, likely on the verge of star formation, and to
determine the initial conditions for low--mass star formation
in the Taurus Molecular Cloud.  The dust observation allows us to infer 
the density structure of the core and together with measurements 
of CO isotopomers gives us the CO depletion.
\NTHP \ and \NTDP \ lines  are good tracers of the dust
continuum and thus they give kinematic information on the core nucleus.  
We derived in this object a molecular hydrogen number
density  n(\HH)$\sim  10^6$\percc \ and a CO depletion factor, 
integrated along the line of sight, 
$f_{\rm D}$ $\equiv$ 9.5$\times$10$^{-5}$/$x_{\rm obs}$(CO) $\sim 15$ in the
central 20\arcsec , similar to the pre--stellar core L1544.  However, 
the $N(\NTDP )/N(\NTHP )$ column density ratio is $\sim$ 0.1, a factor of 
about 2 lower than that found in L1544.  The observed relation between the 
deuterium fractionation and the integrated CO depletion factor across the core 
can be reproduced by chemical models if \NTHP \ is 
slightly (factor of $\sim$2 in fractional abundance) depleted in the central 
3000 AU. The \NTHP \ and \NTDP \ linewidths 
in the core center are $\sim$ 0.3 \kms , significantly
larger than in other more quiescent Taurus starless cores but similar 
to those observed in the center of L1544.  
The kinematical behaviour of L1521F is more complex than 
seen in L1544, and a model of
contraction due to ambipolar diffusion is only 
marginally consistent with the present data.  
Other velocity fields,  perhaps produced by accretion 
of the surrounding material onto the core and/or unresolved substructure, 
are present. Both chemical and kinematical analyses suggest that  
L1521F is less evolved than L1544, but, in analogy with L1544, 
it is approaching the ``critical'' state.  
} \maketitle

\section{Introduction}
   The pre--stellar core L1544 has recently been the subject of
much study because while apparently in hydrostatic equilibrium, there are
indications that it is close to the ``critical'' state at which it will become
gravitationally unstable and from which it will dynamically collapse
(see discussions by Tafalla et al.~\cite{taf98}; Ciolek \& Basu~\cite{cb2000}; 
Caselli et al.~\cite{cas02a}; Caselli et al.~\cite{cas02b}).
   If correct, this is of fundamental importance because it defines the initial 
conditions for the formation of a protostar, which affects many 
theoretical studies of low mass star formation.\\
   Clearly  trying to extrapolate general trends from a single object is difficult 
and a larger number of L1544-like cores (preferably with the same external 
environment) should be studied.
   Unfortunately there are rather few other objects with similar
   properties due to the short timescale of this phase.  
   According to the Ciolek \& Basu (\cite{cb2000}) model, for example, (contraction of a disk 
   driven by ambipolar diffusion) L1544--like  properties fit the model structure of a 
core at times~3-$8\times 10^4$ years prior to the collapse, 
after an evolution of 
$2.6\times10^6$ years. Thus in that particular model, L1544 finds itself in the last 
few percent of its evolution prior to becoming a protostar. 
   While this may be a somewhat too literal interpretation of the
   model results, it shows that ``L1544--type
   cores'' should be relatively rare.

Further progress requires the definition of what is
   a L1544-like core.
  One answer is to use current estimates of dust emission
   and absorption selecting cores of dust extinction upwards of 50 mag. 
   Another approach is to say that cores  which show signs of
   infalling gas (as does L1544, see Williams et al.~\cite{Wil99}; Tafalla et 
al.\cite{taf98}) are ``L1544-twins''.  This latter indicator
   is complicated by the fact that at the high densities found
   in the nuclei of cores similar to L1544, many molecular species and in
   particular CO and CS freeze--out onto dust grain surfaces (see Kramer et 
al.~\cite{kra99};
   Caselli et al.~\cite{cwt99}; Bacmann et al.~\cite{bac02}; Bergin et 
al.~\cite{berg02}; J\o rgensen et al.~\cite{jsv02}; Tafalla et al.~\cite{taf02}); 
observing such tracers implies observing the low density 
surrounding envelope.  However, recent studies indicate that species whose 
abundance is linked to that of molecular nitrogen such as N$_{2}$H$^{+}$ and 
NH$_{3}$ (as well as their deuterated
   counterparts) do not condense out in the same  fashion and hence can be used
   as tracers of the dense gas (Bergin \& Langer~\cite{bl97}).  
The extent to which this is true is
   debatable but it is a useful hypothesis and substantiated by the general
   similarity of the spatial distributions seen for example in dust
   emission and in maps of \NTHP (Tafalla et al.~\cite{taf02}, ~\cite{tmc03}).\\
   
   Caselli et al. (\cite{cas02a},\cite{cas02b}) 
have used \NTHP \ and \NTDP \ to derive the physical, 
chemical and kinematical properties of L1544.
   They found that L1544 has a central  \NTHP ~column density of
   $1.5 \times 10^{13}$ cm$^{-2}$ and a column density ratio $N(\NTDP )/N(\NTHP )$
   of  0.24.
   The \NTHP \ linewidths towards the nucleus (the dust emission peak) are
   roughly 0.3 km s$^{-1}$ and decrease as one goes to positions away from the
   center. The  line of sight velocity measured in \NTHP (1-0) and \NTDP (2-1) shows
   a gradient along the minor axis of the elliptical structure seen in 1.3mm
   dust emission but no clear gradient along the major axis. 
   
   In this paper, we will study another core in the Taurus complex,
 L1521F (at an assumed distance of 140 parsec) using 
   the same approach as in our study of L1544.      
Repeating the L1544 study carried out by Caselli et al. (\cite{cas02a},
\cite{cas02b}) 
is important because it
allows us to check  to what extent L1544 is an exceptional case. 
   In order to do this we need another source which 
  has the same general characteristics as L1544. 
   The source selection was made using some preliminary results we obtained 
in a survey carried out at the IRAM-30m telescope.
   L1521F stood out as being the only core in Taurus, besides L1544, with 
strong \NTDP (2-1)
   emission compared to  \NTHP (1-0).
   This suggests enhanced deuterium fractionation implying an advanced 
   evolutionary state (Caselli et al.~\cite{cas02b}).  Previous observations
of this object have been carried out by Mizuno et al. (\cite{moh94}),
Onishi et al. (\cite{omk96}),  
Codella et al. (\cite{cod97}), and Lee et al. (\cite{lom99}).  
Onishi et al. (\cite{omf99}) also studied L1521F (which they
call MC 27), and found a high central density, suggesting that this 
is the most evolved starless condensation in Taurus. L1521F was also
noted by Lee et al.~(\cite{lmt99}) as a strong infall candidate, in their
survey of CS and \NTHP \ lines in starless cores, although later
mapping of the two tracers  has shown extended ``red'' asymmetry in the 
CS(2--1) profiles (Lee et al.~\cite{lmt01}). 

   In section 2 of this paper, we describe our observational procedure.
   In section 3 we present the observational results deriving
   the physical characteristics of the source and analysing its chemical and 
kinematical properties. In section 4 we discuss the observational results 
and the summary can be found in Section 5. 

\section{ Observations}
    
   The observations were carried out between
April  2002 and January 2003 at the IRAM-30m
   in three different runs.

   In April 2002,
we observed the core in \NTHP (1-0), \NTHP (3-2), \NTDP (2-1) and 
\NTDP (3-2). In general, we used the symmetric frequency switch mode 
and the facility 
  autocorrelator; in Table~\ref{tsettings}, we summarize the main observational
parameters.  The frequencies of the \NTHP (1--0), \NTDP (2--1) and \NTDP 
(3--2) have been updated following the recent determinations of Dore et 
al. (\cite{dcb03}); 
the values in the table refer to the $F_1 \, F$ = 2~3 $\rightarrow$ 
1~2, 2~3 $\rightarrow$ 1~2, and 4~5 $\rightarrow$ 3~4 hyperfine components of 
the \NTHP (1--0), \NTDP (2--1) and \NTDP (3--2) transitions, respectively. 
For \NTHP (3--2) we used the frequency of the 2~1$\rightarrow$1~0 component 
as determined by Caselli et al. (\cite{cas02a}).  
   In the case of the April 2002 \NTDP (3-2) observations,
 in order to improve the baseline
   quality, we used also  
the "Wobbler switching" mode with 
a 240\sec \ throw. 
   We reached an r.m.s. sensitivity in main-beam brightness units of about 100 mK
   in all lines except \NTHP (3-2) ($\sim$ 400 mK).  
   The pointing was checked every 2 hours by means of a 3 or 2 mm continuum 
   scan on nearby quasars and was accurate to within $\sim$ 4\sec .

   In order to refine the maps, originally taken with a 20\sec \ 
spacing, we observed in Nov. 2002 with a 10\arcsec \ grid
(but 5\sec \ spacing in the inner 20 seconds).
\begin{table}[htbp]
\caption{Telescope settings and parameters.}
\begin{tabular}{lccccc}
\hline
line & frequency & HPBW & F$_{throw}$ &T$_{SYS}$ & $\Delta v_{res}$\\
     & GHz	 & \sec &  kHz        & K     & km s$^{-1}$ \\
\hline
\NTHP (1-0)   &  93.1737725 & 26 &  7.5 &  150 & 0.063 \\
\NTHP (3-2)   & 279.511385 &  9 & 14.3 & 1900 & 0.021 \\
\NTDP (2-1)   & 154.217137 & 16 &  7.5 &  370 & 0.038 \\
\NTDP (3-2)   & 231.321966 & 10 & 14.3 &  750 & 0.050 \\
\CEIO (1-0)   & 109.782160 & 22 &  ... &  170 & 0.026 \\
\CEIO (2-1)   & 219.560319 & 11 &  ... &  450 & 0.033 \\
\CSEO (1-0)   & 112.358988 & 21 &  7.5 &  230 & 0.026 \\
\CSEO (2-1)   & 224.714368 & 11 & 14.3 &  730 & 0.052 \\
\hline
\multicolumn{6}{l}{Col. 2 line rest freq., Col.3 Half Power
Beam Width} \\
\multicolumn{6}{l}{Col.4 Freq.Throw, Col. 5 Syst. Temperature (K)
}\\
\multicolumn{6}{l}{Col. 6 Channel Spacing }\\
\label{tsettings}
\end{tabular}
\end{table} 

   Between January 2003 and March 2003, we obtained
continuum data at 1.2mm together with observations of
 \CEIO \ and \CSEO (1--0) and 
(2--1). These data were  taken in service mode by the IRAM staff.

   The continuum data were obtained using MAMBO II, 
the 117-channels bolometer 
available at the 30m.
   We mapped the core within an area of 
150\sec$\times$150\sec \ 
scanning in azimuth  
with a 5\sec/sec speed and an interval between the subscans of 8\sec.
The atmospheric attenuation was measured to be $0.14$
based on tipping curves measured after the map. 
 The data were calibrated using the sources HL Tau and
L1551 for which we assumed fluxes of 0.9 Jy and 1.4 Jy respectively
and the final sensitivity was 5 mJy per 10.5\sec \ beam.   The
calibration error inherent in this comparison is likely to
be at least ten percent due to both atmospheric fluctuations
and calibration errors.

The \CEIO \ data were taken using the on--the--fly technique.
   We simultaneously mapped the \CEIO (1-0) and \CEIO (2-1) using both 
polarizations for each line.
   The  area covered was 150\sec$\times$150\sec \  
and was scanned in the 
Right Ascension direction;
   the distance between the subscans was 5\sec \ as 
was the angular  separation 
between two successive dumps.
   We also obtained  a 9 points map in 
\CSEO (1-0) and \CSEO (2-1) 
   centered 
 at the dust peak and spaced by 20\sec.

\section{Results}
\label{sresults}  

\subsection{Integrated intensity maps and continuum emission}
\label{integrated}
    We show the observed map of the
1.2mm dust continuum in Fig. ~\ref{fig1a}, and 
the map of integrated line intensity, obtained
in the observed transitions of \NTHP , \NTDP , and \CEIO , in 
Fig.~\ref{fig1b}.
   The reference position for these maps is (04:28:39.8, 26:51:35) in
 J2000 coordinates. 

   The bolometer map was reduced with the IRAM standard
reduction program NIC. From Fig.~\ref{fig1a}, we see 
that the observed emission has a ``cometary'' structure in the
sense that the low level contours are well--fitted by an ellipse
with the maximum offset from the center. 
   We can fit the general
elliptical structure of the core with a
2D-gaussian centered at offset
(-30\sec,20\sec) with full width half--power dimensions of 
274\sec $\times$ 170\sec and position angle 25\degr .
If the more extended emission is not included in the fit, 
the 2D--gaussian is centered on the dust peak position
at (-10\sec,0\sec), has half--power dimensions of
127\sec  $\times \, 77 $\sec ,
 a position angle of 25\degr \ and an aspect
ratio equal to 1.6.   The peak intensity is 
90 mJy/beam.  
   
   The spectral line data  were reduced using 
the standard IRAM package CLASS. A summary of line
parameters at the dust peak  is given in Table  ~\ref{linepar} 
and the corresponding spectra are shown in Fig. ~\ref{spectra}.
  The spectra  shown in Fig. ~\ref{spectra}
 (as well as the values in 
Table ~\ref{linepar})  have been derived from data
gaussian--smoothed to a resolution of 26\sec \
in the cases of \NTHP  \ and \NTDP \ ,
but unsmoothed in the case of \CEIO .  These are also the effective
resolutions of the \NTHP \ maps shown in Fig. ~\ref{fig1b} (the 
two \NTDP \ maps are smoothed to 16\arcsec , the angular 
resolution at the (2--1) frequency). 

    One clear  result from Fig. ~\ref{spectra} 
is that in L1521F  there is evidence for an ``infall signature'' 
although less marked than in L1544, 
 where it is attributed to extended infall onto the core
(Williams et al.~\cite{Wil99}) or to central infall in a depleted core 
nucleus (Caselli et al.~\cite{cas02a}).  In fact, we see evidence for 
asymmetric profiles, with the blue peak brighter than the 
red peak, in \NTHP \ (1--0) 
(where we derive an optical depth of  order 4
in the main component based on our fit to the hyperfine satellites),
but the two peaks are not clearly separated as in the case of L1544. 

It is interesting to note that two peaks are
 present in the spectrum of \NTHP (3--2) line
toward the offset (-10,10), as shown in Fig.~\ref{n2h+32}.  
 The hyperfine structure (hfs) 
fit to the \NTHP (3--2) line, assuming the presence of two 
velocity components along the line of sight,  
gives $V_{\rm LSR \, 1}$ = 6.35$\pm$0.1 \kms \ and  $V_{\rm LSR \, 2}$ 
= 6.55$\pm$0.04 \kms .  The line
widths of the two components are $\Delta v_1$ = 0.15$\pm$0.02 \kms \
and  $\Delta v_2$ = 0.19$\pm$0.06 \kms , marginally (factor $\la$ 1.5) 
broader than the thermal \NTHP \ line width at 10 K.  Although this 
result should be confirmed with higher sensitivity data, the velocities
are consistent with those of the N1 and N2 (\NTHP ) clumps observed by 
Shinnaga et al. (\cite{sol03}) with interferometric observations.  The hfs fit 
to the \NTHP (1--0) line,  
assuming two velocity components and fixing the velocities 
to the values found with the (3--2) line, is consistent with the observed 
spectrum (if the two velocities are not fixed, the hfs fitting procedure 
applied to the \NTHP (1--0) spectrum does not converge, probably because
the two components are not well separated).  
This suggests that the two clumps observed by Shinnaga et al. (\cite{sol03})
in \NTHP (1--0) are also present in our single dish data, although
the \NTHP (1--0) extended emission partially hides the features arising 
from the central region.

\begin{table*}[htbp]
\caption{Line parameters derived from hyperfine structure fitting 
at position (-10,0). The values refer to spatially
averaged spectra  (26\sec beam).}
\begin{tabular}{lccccc}
\hline
line       & V$_{LSR}$   & $\Delta$V  & $\tau$ & $\int T_{\rm{MB}}dV ^b$& 
T$_{\rm ex}$ \\
           & km s$^{-1}$ & km s$^{-1}$&        &  K km s$^{-1}$      &  K     \\
\hline
\hline
\NTHP(1-0) & 6.472$\pm$0.001&0.299$\pm$0.001&17.9$\pm$0.1&5.859$\pm$0.016&4.94$\pm$0.03 \\
\NTHP(3-2) & 6.388$\pm$0.010$^a$&0.290$\pm$0.030& 1.4$\pm$1.0&0.692$\pm$0.075&5.1$\pm$3.7 \\
\NTDP(2-1) & 6.505$\pm$0.004&0.268$\pm$0.010& 2.2$\pm$0.4&0.978$\pm$0.026&4.6$\pm$0.9 \\
\NTDP(3-2) & 6.507$\pm$0.006&0.222$\pm$0.024& 1.2$\pm$0.9&0.332$\pm$0.029&4.1$\pm$3.2 \\
\hline
\multicolumn{6}{l}{NOTE: $^a$ The \NTHP (3--2) frequency is as in Caselli et 
al. (\cite{cas02a},}\\
\multicolumn{6}{l}{whereas for the other lines we used the updated values 
determined by}\\
\multicolumn{6}{l}{Dore et al. (\cite{dcb03}). $^b$ The integration includes all hyperfines.} 
\end{tabular}
\label{linepar}
\end{table*}

  Fig. ~\ref{fig1b} shows that L1521F has in common with 
several other cores   
    (see also Tafalla et al.~\cite{taf02};
Caselli et al.~\cite{cas02a},\cite{cas02b}) the property that while the maps of 
   the nitrogen bearing molecules have  a similar
appearance to those observed in dust emission, maps
in rare CO isotopomers such as \CEIO \ show no correlation
with the dust.  In fact, the distribution of \CEIO (1--0)
and (2--1)  is essentially flat over the area we have mapped, suggesting
perhaps that the layer sampled in \CEIO \ is either in the
background or foreground relative to that seen in dust emission.

 We have been able to make some estimate of the optical depths  of
the observed \CEIO \ lines by comparison with our 9-point \CSEO
\ map (20\sec \ spacing centred on (-5,-5)).
  From the integrated intensity ratio of
the \CEIO \ and \CSEO (1--0) lines and assuming
an intrinsic abundance ratio [\CEIO]/[\CSEO] of
3.65 (from [$^{18}$O]/[$^{17}$O] = 3.65, Penzias~\cite{p81};
see Kramer et al.~\cite{kra99} for a discussion of the validity
of these techniques), we  can put in general upper limits on
the optical depth of \CEIO (1--0) of order unity. At the (-5,-5)
and (-25,-5) offsets, there is however 
evidence that \CEIO (1--0) is optically thick  and in fact
we derive  \CEIO (1--0)  optical depths of 1.0 and 1.5
respectively
at these two offsets (errors 50\% ). In these positions
we used the \CSEO (1--0) data and, assuming 
$T_{\rm ex}$ = 10~K (consistent with the observed \CSEO (2--1)/(1--0) 
line ratio) and [$^{18}$O]/[$^{17}$O] = 3.65,
we found an average \CEIO \ column density of $\sim$ 2$\times$10$^{15}$
cm$^{-2}$ in the central 25\sec . 

It is also interesting that we find that \NTHP (1--0) and
dust emission maps have similar sizes, whereas \NTDP \ maps 
are somewhat smaller.   
 Fitting a 2D gaussian to
the \NTHP \ and \NTDP \ maps in the same fashion as for
the dust emission, we find the parameters  given
in Table~\ref{gauss2d}  for the angular sizes of \NTHP \
and \NTDP \ in L1521F.  We deduce from these data
a linear size for L1521F seen in \NTHP (1-0) of roughly
14000 AU and dimensions in \NTDP (2-1) of 13000 $\times $
6000 AU. It is interesting to note the smaller sizes
derived in the higher J transitions of both species, suggesting
that they are sampling somewhat higher density gas. 
The 1.2mm dust continuum on the other hand is more 
extended suggesting that either an increase in excitation or in
abundances is causing the \NTDP \ and \NTHP (3--2) line emission to increase
in relative strength toward the dust peak.  Notable also are 
the large aspect ratios (1.6--1.9) observed in \NTDP \ suggesting 
an ellipsoidal or triaxial form for the high density core.
   
\begin{table*}[htbp]
\caption{Angular Dimensions derived from two dimensionsal
gauss fits to maps in \NTHP \ and \NTDP}
\begin{tabular}{lcccccc}
\hline
line & Int. & $\Delta\alpha$& $\Delta\delta$ & Maj. Axis &Min. Axis & P.A.\\
     & \kks	      & \sec            & \sec           & \sec        & \sec  & \deg\\ 
\hline
1.2mm  &   & -8  & 2 &  127  &  77  &  -25 \\
\NTHP (1-0)   &  5.82 & -12 & 1 & 110 & 90 &	-18 \\
\NTHP (3-2)   &  0.68 &  -7 & 1 &  90 & 67 &   +30 \\
\NTDP (2-1)   &  1.02 &  -9 & 4 &  90 & 46 &	 -9 \\
\NTDP (3-2)   &  0.34 & -11 & 5 &  63 & 38 &	-33 \\
\hline
\multicolumn{6}{l}{Col. 2 Integrated intensity at peak, Col.3 R.A. offset of peak} \\
\multicolumn{6}{l}{Col.4 DEC offset of peak, Col. 5 Major axis size (FWHM)}\\
\multicolumn{6}{l}{Col. 6 Minor axis size (FWHM), Col. 7 Position angle,}\\ 
\multicolumn{6}{l}{measured East of North.}\\
\end{tabular}
\label{gauss2d}
\end{table*} 

\subsection{Density and mass distribution}
\label{sdensity}

   We expect the observed continuum emission at mm
wavelengths to be optically thin and can relate the 1.2mm flux
   to the \HH \ column density under
   the approximation of constant dust emissivity and temperature:
 \begin{displaymath}
   N(\MOLH ) = \frac{S_{1.2mm}}{\Omega_{\rm beam} \, B_{\nu}(T)\,
 \kappa_{1.2mm} \; m } ,
 \end{displaymath}
where $N(\MOLH )$ is the averaged \HH \ column density , $S_{1.2mm}$ 
is 
the emitted 1.2mm flux density from the core, $\Omega_{\rm beam}$ 
is the telescope beam solid angle,
 $\kappa_{1.2mm}$ is
the dust absorption coefficient per gram of gas
 at 1.2mm, $B_{\nu}(T)$ is the Planck function
at temperature $T$,  and $m$ the mean molecular mass.  In our
calculations we used a dust temperature $T_{\rm dust}$ = 10 K,  close to the gas 
temperature found by Codella et al. (\cite{cod97}) using ammonia data (9.1 K),
a dust opacity $\kappa_{1.2mm} \, =$
 0.005 cm$^{2}$ g$^{-1}$ (Andr\'e et al. \cite{awm96})
 and $m$ = 2.33 amu (with uncertainties of typically 
50\%; see e.g. Bianchi et al.~\cite{bga03} and Gon\c calves et 
al.~\cite{ggw04}).  
This expression can be integrated over the
1.2mm map to derive the total gas mass. 

 In this fashion, we derive a core mass from the continuum
 measurements, within the large ellipse in Fig.~\ref{fig1a}, 
of 5.5$\pm$0.5 \solmass \  corresponding to an integrated 
total flux of  $3.0\pm 0.3$ Jy. (The error is   dominated by
errors in calibration and baseline removal.)  Within the 
26\sec \ beam, used for our line measurements, the
enclosed mass is 0.7\solmass .  It is interesting to note that 
5\solmass \ and 0.8\solmass \ are the average total mass and 1 Jeans
masss in the Taurus Molcular Cloud complex (Goodwin et al.~\cite{gww04}).
   
   For the purpose of comparison with model calculations, we have
used the 1.2 mm continuum data to estimate the density distribution 
under the assumption of spherical symmetry.  This inevitably involves a
rough approximation since the L1521F core clearly {\it is not spherically
symmetric} (the aspect ratio in the 1.2mm continuum emission is 1.6, see 
Tab.~\ref{gauss2d}) and would be better approximated with an ellipsoid.

 Nevertheless, we have followed the technique adopted by 
Tafalla et al. (\cite{taf02}) and fit our data with a model of 
the form:
   $$ n_{\MOLH }(r) = \frac{n_{0}}{1 + \left( \frac{r}{r_0}
\right)^\alpha} ,$$ 
We thus  circularly averaged
the 1.2 mm data around the peak and used a $\chi^2$  routine to
fit the dust profile using the above model integrated along the line
of sight and convolved with a 10.5\sec \ gaussian.
   The result of the fitting procedure can be seen in Fig.~\ref{dprof} . The
   parameters that best fit our data are $n_{0}= 10^6$
   cm$^{-3}$, $r_0 = 20$\sec \ (corresponding
to 0.014 parsec or 2800 AU) and $\alpha = 2.0$.

\subsection{CO freeze--out}
\label{scofreeze}

The comparison between the dust continuum emission and the 
\CEIO \ integrated intensity map allows the determination of the 
amount of CO freeze--out onto dust grain surfaces, integrated along
the line of sight.  This is possible because 
the millimeter continuum data furnish $N(\MOLH )_{\rm 1.2mm}$, 
the column density of 
molecular hydrogen across the core, assuming optically thin conditions.  
The same quantity  is obtained from \CEIO \ lines ($N(\MOLH )_{\rm CO}$), 
 again under the assumption
of optically thin conditions, and adopting the relation between CO and 
\MOLH \ valid in undepleted conditions ([CO]/[\MOLH ] $\simeq$ 
9.5$\times$10$^{-5}$ $\equiv$ $x({\rm CO})_{\rm can}$, 
the ``canonical'' CO abundance value; Frerking et al. \cite{flw82}).  From the 
$N(\MOLH )_{\rm 1.2mm}$/$N(\MOLH )_{\rm CO}$ column 
density ratio, the integrated CO depletion factor ($f_{\rm D}$), or the 
ratio between the canonical and the observed CO abundance 
($x({\rm CO})_{\rm can}$/$x({\rm CO})_{\rm obs}$), is easily derived.

In practice, this process requires the division of the 1.2mm dust continuum
emission (${\cal F}_{\rm 1.2mm}$[mJy/beam] = 
$S_{\rm 1.2mm}$[mJy]/$\Omega_{\rm beam}$) 
map by the \CEIO \ integrated intensity 
($W_{\CEIO}$) map. The map--division has been carried out using
the IRAM image manipulation software GRAPHIC, 
after degrading the continuum map to the angular resolution of the 
\CEIO (1--0) observations (22\sec ) so that the integrated depletion factor 
can be expressed by:
\begin{eqnarray}
f_{\rm D} & = & x({\rm CO})_{\rm can} \times 
	\frac{N(\MOLH )_{\rm 1.2mm}}{N(\CEIO )} 
	\frac{[\rm ^{18}O]}{[\rm ^{16}O]} \nonumber \\
          & = & 8.5\times 10^{-2} 
	\frac{{\cal F}_{{\rm 1.2mm}}({\rm mJy/22\sec beam})}
	{W_{\CEIO}({\rm K \, km \, s^{-1}})} 
	 \label{efd}.
\end{eqnarray}
To derive eq.~(\ref{efd}) 
we have assumed a dust temperature $T_{\rm dust}$ = 10 K,  
$\kappa_{\rm 1.2mm}$ = 0.005 cm$^2$ g$^{-1}$, and an abundance ratio 
$[\rm ^{16}O]/[\rm ^{18}O]$ = 560 (Wilson \& Rood \cite{wr94}). 

 There are several {\it caveats} to the above procedure.  One is
 that the ``canonical abundance'' appears to vary from cloud to
cloud (Lacy et al. \cite{lkg94}; 
Alves et al. \cite{all99}; Kramer et al.~\cite{kra99}) 
and is roughly one third of the diffuse cloud carbon gas 
phase abundance (Sofia et al. \cite{scg97}).  Given that CO
is expected to represent essentially all the gas phase carbon in
molecular clouds, this suggests depletion of carbon in some 
form  (not necessarily as solid CO) even  on the outskirts of cores
(we note that the direct study of CO solid state
features in Taurus (Chiar et al. \cite{cak95}) shows 
that solid CO towards 4 field stars in Taurus is less  than
40\% \ of the canonical gas phase CO abundance  and is
observed for extinctions $A_{\rm V}$ greater than 6 magnitudes). 
 Thus we conclude that in particular cases such as L1521F,
it is quite possible that we are using a value of 
$x({\rm CO})_{\rm can}$  which is a factor of order 2 too large or 
small thus influencing the values of $f_{\rm D}$ which we derive
but not the trends over our map. 

 Another  problem  is that the values of $f_{\rm D}$ which we
derive are integrated along the line of sight in a situation
where the observed \CEIO \ emission forms in an outer shell 
whereas the dust emission  mainly emanates from the dense
core nucleus.  As a consequence, we observe \CEIO \ mainly
from foreground and background gas which (see Fig.~\ref{fig1b}) has
an essentially flat distribution over the region mapped by
us. One concludes that our values of $f_{\rm D}$ are strict
lower limits to the CO depletion {\it in the core nucleus} 
from which
dust emission is observed.   It is also the case that in
this situation, the form of our map of $f_{\rm D}$ is essentially
that of the dust emission (as we indeed find, see eq.~\ref{efdh2}). 
The {\it local} distribution of the CO depletion factor 
(which we call ${\cal F}_{\rm D}$, see below) may differ greatly
and be much more highly peaked than in our map. 
 Nevertheless, the  $f_{\rm D}$   values derived by us are
direct observables and we have therefore used them for the
purpose of correlating with parameters such as the observed
deuterium fractionation.  We have also used them for
model comparisons (sections~\ref{ssimple} and ~\ref{sconcentrated}) 
bearing in mind the above discussion. 

In Fig.~\ref{fd_dust} the $f_{\rm D}$ map is shown (thick contour)
overlapped with the smoothed 1.2mm map (grey scale) and the \NTDP (2--1) 
map (dashed contours).
The CO depletion factor increases between 6 at the lowest contour of
the 1.2mm map (${\cal F}_{\rm 1.2mm}$ = 
95 mJy/22\arcsec \ beam) to 18 at the peak
position (offset [0,9]), which is 11\arcsec \ off   
the millimeter dust emission peak (offset [-8,+2]), where 
${\cal F}_{\rm 1.2mm}$ = 
320 mJy/22\arcsec \ beam).  From Fig.~\ref{fd_dust} we immediately see
that $f_{\rm D}$ correlates with the \NTDP \ emission (and deuterium
fractionation, see Sect.~\ref{sdeut}) and the 1.2~mm dust flux. 
The good correlation between 
$f_{\rm D}$ and ${\cal F}_{\rm 1.2mm}$ is also evident in Fig.~\ref{fd_cd}, 
where $f_{\rm D}$ is 
plotted as a function of $N(\MOLH )_{\rm 1.2mm}$ 
($\equiv$ 4.25$\times$10$^{20} {\cal F}_{\rm 1.2mm}$ 
mJy/(22\arcsec \ beam), with the assumptions on $T_{\rm dust}$ and  
$\kappa_{\rm 1.2mm}$ as described above).  Thus, in L1521F, and with 
the caveats discussed above, $f_{\rm D}$
is linearly dependent on the \MOLH \ column density ($N(\MOLH )$);
the best fit to the data in Fig.~\ref{fd_cd} (dashed curve) 
gives:   
\begin{eqnarray}
f_{\rm D} & = & 1.5 \, 
\left[ \frac{N(\MOLH )}{10^{22} {\rm cm}^{-2}} \right]^{0.9} .
\label{efdh2}
\end{eqnarray}
In retrospect, this is not surprising as $N(\CEIO )$ does not vary
greatly over the map. We note that $f_{\rm D}$ is always $\ga$ 2
 in the region traced by the present observations,
suggesting that even at the core edges (where $N(\MOLH )$ 
$\sim$ 10$^{22}$ cm$^{-2}$) CO molecules are partially 
($\simeq$ 30\%) frozen onto grain mantles. This result is  
consistent with the average value of CO depletion ($\simeq$ 25\%) gauged from 
observations of solid CO features in the direction 
of background stars in the Taurus complex (Chiar et al. \cite{cak95}).
The thick curve in Fig.~\ref{fd_cd} is the result of a 
simple chemical model of a centrally concentrated cloud, where 
CO depletion is taken into account, and where the CO abundance
has been integrated along the line of sight to obtain the observed
$f_{\rm D}$ ($f_{\rm D}$ = $\int {F}_{\rm D}(r) \, dl / \int dl$, 
with $F_{\rm D}(r)$
being the CO depletion factor {\it within} the core, thus function 
of the cloud radius $r$; 
see Section~\ref{schemistry}).  We anticipate here that
the $f_{\rm D}$ value at the cloud center is only a very small fraction 
($\sim$ 1\%) of $F_{\rm D}$ in the central few thousand AU of the core.

\subsection{\NTHP -- \NTDP \ column densities, deuterium fractionation, 
and volume densities}
\label{sdeut}

The \NTHP \ and \NTDP \ column densities have been determined using 
the ``constant--$T_{\rm ex}$'' (CTEX) approximation, which reduces to  
simple analytic expressions (see Appendix A in Caselli et al.~\cite{cas02b}), 
and the Large Velocity Gradient (LVG) approximation.  
Both approaches give reasonable column  density estimates  as long
 as  optical depths are small.  When, as for example for \NTHP (1--0),
 the optical depths in the main hyperfine satellites are large, one
 is best (independent of method) to use the weakest of the satellites
 or alternatively the optical depth inferred from the intensity ratio
 of the weakest satellite to the strong main components.  The errors in
 any case  stem from the difficulties in estimating the optical depth
 of thick lines  compounded with possible non-LTE effects for the 
hyperfine satellites (Caselli et al. \cite{cmt95}).
 Errors due to the estimate of the
 partition function (i.e. the fraction of the species in unobserved
 levels) appear to be less.  We  in general report column
 densities for \NTHP \ using the CTEX approach based on the
 integrated intensity of the  weakest satellite of \NTHP (1--0)
 and for \NTDP \ assuming optically thin conditions.  From comparison
 between the different approaches employed by us,
 we estimate the column density errors to be
 30 percent.

  We have also used LVG estimates to infer the density towards the
   peak and edges of our \NTHP \ map. Here, we assume a temperature of
   10 K and have used rates from Green (\cite{g75}) for
   collisions of He with \NTHP .   Based on the values in Tab.~\ref{linepar}
   (data smoothed to 26\arcsec \ spatial resolution) 
 for the integrated intensities of \NTHP , we find
   $n(\MOLH )$ = $4 \times 10^5$ \percc \ towards the dust peak of
   L1521F and 2.5$\times$10$^5$ \percc \ 30\arcsec \ North (offset 
[-10, 30]). For \NTDP (2--1) and (3--2), the corresponding numbers
   are 6$\times$10$^5$ \percc \ and 3.5$\times$10$^5$ \percc , at the peak and 
(-10, 30) offset positions, respectively.  
These values are similar to the corresponding
   values for L~1544 consistent with the idea that they have
   similar density distributions. The density estimates are
   somewhat smaller than estimates based on the dust emission,
probably due to the different (factor of 2.4 lower) spatial 
resolution\footnote{The central density obtained with the continuum data 
reduces to $\sim$ 6$\times$10$^5$ \percc \ when averaged within 26\arcsec.}, 
and consistent perhaps with the idea that \NTHP \ is abundant in
   a shell outside but close to the dust peak.  However, given that 
the LVG method 
assumes homogeneous conditions, the LVG--derived
densities are averages along the line of sight, thus lower values are
expected when compared to the continuum data analysis, which takes into 
account the core density structure.  

The deuterium fractionation is directly estimated from the 
$N(\NTDP )/N(\NTHP )$ column density ratio ($\equiv$ $R_{\rm deut}$), 
and the $R_{\rm deut}$ map in L1521F, assuming CTEX conditions,
 is shown in Fig.~\ref{Dfrac_map}.
$R_{\rm deut}$ ranges between $\sim$ 0.02 at the core edge to 0.1 in 
a region about 20\arcsec \ in size and centered on the dust peak 
position. The peak value of 
$R_{\rm deut}$ is about a factor of 2 smaller than that found in L1544 
(Caselli et al.~\cite{cas02b}). 
We note that the column density of \NTHP \ is similar in the two 
cores, and that the factor of 2 of difference in deuterium fractionation
is due to the (factor of 2) larger \NTDP \ column density in L1544. 
This suggests that, although the two cores are similar in structure, 
L1521F is probably slightly less evolved than L1544 (see 
Sect.~\ref{sdiscussion}). 

%

\subsection{Line width vs. impact parameter}
\label{sdv-b}

Quiescent starless cores mapped in \AMM (1,1) and \NTHP (1--0) lines 
typically show a ``velocity coherent'' zone of nearly constant line width 
($\Delta v$) within the half-maximum contour map, followed by a $\Delta v$ 
rise at larger distances from core center (e.g. Goodman et al. 
\cite{gbw98}).
There are however exceptions to this general trend, as pointed out 
by Caselli et al. (\cite{cbm02}). In particular, L1544 shows a significant
increase of \NTHP \ and \NTDP \ (but not H$^{13}$CO$^+$  and DCO$^+$ ) 
line widths toward the center 
(factor of 1.5 within $\sim$ 50\arcsec ; Caselli et al.~\cite{cas02a}).
This increase has been interpreted as evidence of infall 
in the central few thousands AU, where CO and related species
(such as H$^{13}$CO$^+$  and DCO$^+$) are heavily depleted.  Indeed, 
the line--width increase is consistent 
with models of ambipolar diffusion (see 
Sect.~\ref{skinematics}).

The same trend has been observed in L1521F using \NTHP \ and \NTDP \
lines, and in Fig.~\ref{dv_b} we show the results 
obtained in \NTHP (1--0) lines (the linewidth corrected for optical 
depth, or the intrinsic linewidth, is plotted).  
The figure also shows two 
theoretical predictions which will be discussed in Sect.~\ref{skinematics}.
The decrease of the intrinsic line width with impact parameter in L1521F, 
although not as steep as in L1544, is clear in Fig.~\ref{dv_b}, where
the average value of $\Delta v$ within bins of 15\arcsec \ (see big
dots) ranges between 0.3 \kms \ at the dust peak to 0.25 \kms \ at 
a projected distance of 80\arcsec . 

One might interpret this line
width gradient as being due to increased optical depth towards the dust
peak.  This however seems unlikely as illustrated in Fig.~\ref{dv_phil}, 
where the profiles of the isolated hyperfine component of the 
\NTHP (1--0) line along the 45\deg \ and -45\deg \ axes, passing through 
the dust peak position, are shown.  If the line is self--absorbed toward
the core center, the hyperfine components with the largest statistical 
weight will be broadened compared to the weakest ones, affecting the 
hfs fit. However, we performed gaussian fits to all the hyperfines
components, 
finding the same $\Delta v$ values within the uncertainties.  In fact,
 a similar trend is
also observed if one plots the linewidth of the $F_1 \, F$ = 1 0 
$\rightarrow$ 1 1 (or weakest) component of the \NTHP (1--0) line
versus the impact parameter. This line, being thin and symmetric
across the core, is not affected by self--absorption.

We believe that this is a characteristic of unstable or supercritical 
(e.g. Mouschovias \& Spitzer \cite{ms76}) cores 
on the verge of star formation, more briefly called {\it pre--stellar cores}, 
a term which we use to characterize the {\it subset of starless cores
undergoing central infall} (see also 
Ward--Thompson et al. \cite{wka03}). 

\subsection{Velocity field}
\label{svel_field}

Assuming that L1521F is in solid body rotation, we determined the 
magnitude ${\cal G}$ and the direction $\Theta$ 
of the corresponding velocity gradient 
following the procedure described in Goodman et al. (\cite{gbf93}).  
The magnitude 
${\cal G}$ ranges between 0.4 and 1 km s$^{-1}$ pc$^{-1}$ 
depending on the tracer used, and 
large variations are also obtained for the gradient direction $\Theta$
(see Tab.~\ref{tgradient}, columns 2 and 3).  There is no tendency for 
${\cal G}$ to increase for higher density tracers, as observed in 
L1544 (Caselli et al.~\cite{cas02a}).   

\begin{table}[htbp]
\caption{Results of velocity gradient fits.}
\begin{tabular}{lccc}
\hline
line & ${\cal G}$ & $\Theta ^{a}$ & $<{\cal G}_{\rm l}> ^{b}$  \\
     & (km/s/pc) & (deg) & (km/s/pc) \\
\hline
\NTHP (1-0)   & 0.366$\pm$0.005 & $-$133.6$\pm$0.7 & 1.9$\pm$0.1 \\
\NTHP (3-2)   & 1.1$\pm$0.1 & $-$168$\pm$9 & 2.7$\pm$0.3  \\
\NTDP (2-1)   & 0.95$\pm$0.06 & $+$145$\pm$5 & 3.5$\pm$0.2 \\
\NTDP (3-2)   & 0.4$\pm$0.2 & $-$128$\pm$27 & 4.7$\pm$0.4  \\
\hline
\multicolumn{4}{l}{NOTE: $^{a}$ Direction of increasing velocity, measured East} \\
\multicolumn{4}{l}{of North. $^{b}$ Mean values of the magnitude of local 
 velocity} \\ 
\multicolumn{4}{l}{gradients and corresponding standard error.} \\
\label{tgradient}
\end{tabular}
\end{table} 

To investigate in more detail the internal motions of L1521F, we 
determined  ``local'' velocity gradients (see Caselli et al.~\cite{cas02a}), where
${\cal G}_{\rm l}$ and $\Theta_{\rm l}$ have been calculated in adjacent 
3$\times$3--point grids of the maps. The results are shown
in Figs.~\ref{fhgrad} and ~\ref{fdgrad} for \NTHP \ and \NTDP \ maps, 
respectively. The arrows across the map indicate the magnitude and the direction 
of local gradients, and they are centered on the 9--point grid used to 
estimate the corresponding ${\cal G}_{\rm l}$ and $\Theta_{\rm l}$
values. From the figures it is clear that L1521F is {\it not} 
undergoing solid body
rotation.  The velocity structure is quite complex, showing 
portions of the core where the gradient direction changes rapidly.
For example, if we concentrate on the 
\NTHP (1--0) map (see Fig.~\ref{fhgrad} [top panel]), 
which has the highest sensitivity, three converging 
velocity gradient patterns are clearly 
visible: (i) toward South--West in the Northern half, (ii) toward 
North--East in the SE quadrant, and (iii) toward North--West in the 
SW quadrant of the core.  A similar structure is also present in the
other maps (see Figs.~\ref{fhgrad} [bottom panel] and ~\ref{fdgrad}).

Interestingly, the mean of the local--gradient magnitudes ($<{\cal G}_{\rm l}>$)
increases going from \NTHP (1--0) to \NTDP (3--2) (see column 4 of 
Tab.~\ref{tgradient}), suggesting that internal motions, although complex
($\Theta_{\rm l}$ widely varies across the cloud),
tend to become more prominent toward the core center. Moreover, \NTDP \ 
gradients appear larger than those derived from \NTHP . If the observed
velocity structure is at least partially due to inward motions in the core 
center, the larger local--gradient magnitudes detected in \NTDP \ can be  
explained by \NTDP \ being a better tracer of the core central regions, as found
in L1544 (Caselli et al.~\cite{cas02a},b; see also Sect.~\ref{schemistry}).  
However, the magnitude of the local gradients is 
generally within a few km s$^{-1}$ pc$^{-1}$ (see Tab.~\ref{tgradient}), thus 
they are produced by small velocity variations ($\delta v \simeq $ 0.02--0.05 \kms ) 
within $\simeq$ 0.01 pc, the size of the grid where local gradients 
have been estimated (see also Sect.~\ref{skinematics} for discussion).

We note that the velocity gradient found with the present \NTHP (1--0)
observations is significantly smaller than that deduced by Shinnaga et al. 
(\cite{sol03}) using interferometric observations, probably because they resolve
out the more extended emission, with smaller or almost opposite (see 
Shinnaga et al.~\cite{sol03}) 
velocity gradients, compared to the central regions.

\section{Discussion}
\label{sdiscussion}

The previous sections described the results found in our analysis of L1521F.
The density profile, the CO depletion factor, the deuterium fractionation 
and the velocity field in the core have been presented.  We found several 
similarities between L1521F and L1544, including the density structure (with
$n(\MOLH )$ $\simeq$ 10$^6$ \percc \ at the center), 
the amount of integrated 
CO depletion ($f_{\rm D}$ $\simeq$ 15) toward the dust peak,
and the decrease of line width with increasing distance from the cloud center.
Differences are however present in the amount of deuterium 
fractionation (factor of $\sim$ 2 less than in L1544) and in the velocity 
structure.
In this section we will discuss these findings in view of our knowledge of 
the chemical and physical processes in dense cloud cores.  The discussion is 
thus split in two subsections, one focussing on the chemistry and the other
on the kinematics of L1521F.  

\subsection{Chemistry}
\label{schemistry}

In cold clouds, the exothermicity of the proton--deuteron exchange reaction 
\begin{eqnarray}
\HTHP \ \, + \, {\rm HD} \, & \rightarrow & \, \HTDP \ \, + \HH + \Delta E ,
\label{eh2dp}
\end{eqnarray}
where $\Delta E/k$ = 230 K (e.g. Millar et al. 
\cite{mbh89}), is responsible for the 
enhancement of the \HTDP /\HTHP \ abundance ratio well above the local
elemental abundance of deuterium (1.5$\times$10$^{-5}$; Oliveira et al.
\cite{ohk03}). 
As a consequence, given that \HTDP \ transfers its deuterium to neutral 
species such as CO and \MOLN , producing \DCOP \ and 
\NTDP , the $N(\DCOP )/N(\HCOP )$ and $N(\NTDP )/N(\NTHP )$ 
column density ratios reach the values observed towards dense cloud cores 
($\simeq$ 0.02--0.2; e.g. Butner et al. \cite{bll95}, Williams et al. 
\cite{wbc98}, Caselli et 
al.~\cite{cas02b}, Sect.~\ref{sdeut}).  

However, the freeze--out of neutral species, 
such as CO, O, and \MOLN ,  boosts the deuterium fractionation  
(e.g. Dalgarno \& Lepp \cite{dl84}).  In fact, a decrease in the fractional 
abundance of gas phase CO leads to a decrease of the \HTHP \ and \HTDP \ 
destruction rates and to an increase (caused by the higher \HTHP \ 
abundance) of the \HTDP \ formation rate
 (e.g. Roberts \& Millar \cite{rm00a}; see reaction~\ref{eh2dp}). 
An empirical relation between CO depletion and deuterium fractionation in 
prestellar cores has been determined by Bacmann et al. (\cite{blc03}) 
using doubly deuterated 
formaldehyde, whose abundance is also predicted to largely increase with CO 
freeze--out (Roberts \& Millar \cite{rm00b}; see also Roberts et al. 
\cite{rhm03}). 

In the previous sections, we found that in L1521F, CO is depleted 
(with percentages ranging from 30\% at the core edge to 93\% at the 
center, deduced from the integrated CO depletion factor) and, 
similarly to L1544, the deuterium fractionation is large. 
The present estimates of $R_{\rm deut}$ $\equiv$ $N(\NTDP )/N(\NTHP )$ as a 
function of distance from the dust peak allow us to study the relation between 
deuterium fractionation and CO freeze--out across a dense core for the 
first time and test chemical models. 
In Fig.~\ref{rd_fd}, $R_{\rm deut}$ is plotted as a function 
of $f_{\rm D}$ for all the positions present in Fig.~\ref{Dfrac_map}.  
We note a clear tendency for the deuterium fractionation to increase
with integrated CO depletion factor.  

In the following, we will analyse the 
observational results with simple chemical models, to better understand
the observed $f_{\rm D}$--$N(\MOLH )$ and $R_{\rm deut}$--$f_{\rm D}$ trends 
shown in Figs.~\ref{fd_cd} and \ref{rd_fd}, and the differences between
L1521F and L1544. It will be interesting to see whether in L1521F there is
evidence of the so--called ``molecular hole'', or the region where
all species heavier than helium are heavily ($\ga$ 98\%) depleted from
the gas phase, as deduced in the case of L1544. In fact, the recent detection
of \HTDP \ toward the L1544 dust peak is consistent with the 
presence of a molecular hole in the central 
$\sim$ 2800 AU (Caselli et al. \cite{cvc03};
Walmsley et al. \cite{wfp03}). Then, the \NTHP \ and \NTDP \ emission 
maps should show a central hole or emission plateau 
of similar size, but this has not been observed perhaps because of the 
poor spatial coverage of the central region and the limited 
spatial resolution ($\sim$ 2000 AU; Caselli et al.~\cite{cas02a}). 
Such a ``molecular hole'' was predicted by Caselli et al. (\cite{cwt99}) 
and Caselli et al. (\cite{cas02a})
 in an attempt to interpret the origin of double peaked
optically thin lines for the case of L1544, 
including the thinnest \NTHP (1--0) hyperfine component.
In the case of L1521F, molecular lines, although asymmetric, 
do not clearly show two separated peaks.  Thus, if the line asymmetry is 
due to the presence of a molecular hole in a contracting core, 
the hole should have a smaller size than in the case of L1544.  Evidence for 
\NTHP \ depletion toward core centers has also been claimed in B68 
(Bergin et al.\cite{berg02}) and L1512 (Lee et al. \cite{les03}), two starless cores 
with  central densities $\sim$ 10$^5$ \percc \ and close to hydrostatic 
equilibrium, thus in a different chemical and dynamical phase 
compared to L1544 and L1521F. 

\subsubsection{A simple analytical model}
\label{ssimple}

The four thin curves in Fig.~\ref{rd_fd} refer to the outputs of a 
simple 
steady state analytical chemical model including \HTHP , \HTDP , \HH , HD, CO, 
\MOLN , \HCOP , \DCOP , \NTHP , \NTDP , 
electrons and negatively charged grains.  The recombination
on grain surfaces uses the rates from Draine \& Sutin (\cite{ds87}), 
assuming that 
the grains are bare and that their
abundance by number is given by the MRN (Mathis et al. 
\cite{mrn77}) distribution 
with upper cutoff radius of 0.25 $\mu$m and lower cutoff radius 
$a_{\rm min}$ = 50 \AA (standard case, solid curves), and that all 
grains are negatively charged (a more realistic value for the 
fraction of negatively charged grains may be $\sim$ 0.5; see 
Flower \& Pineau des For\^ets~\cite{fp03}).  We also 
considered a larger  $a_{\rm min}$ (= 500 \AA , dashed curves) to 
roughly take into account the process of grain coagulation in 
dense cores (e.g. Ossenkopf \& Henning 
\cite{oh94}).  As expected, {\it larger grains 
cause a (slightly) higher deuterium fractionation}, given that the number
density of dust grains decreases and does also the recombination rate 
on grains of \HTDP \ molecular ions (further increasing the grain size does 
not significantly change the result, given that dissociative 
recombination becomes more important).  This simple model also 
assumes that the electron fractional abundance $x(e)$ varies with density
as 1.3$\times$10$^{-5} \, n(\HH )^{-0.5}$ (McKee \cite{mk89})\footnote{The 
use of the relation found in L1544 by Caselli et al. (\cite{cas02b}) 
shifts the theoretical
curves up by a factor of $\sim$ 1.3 in deuterium fractionation.},  
and that $n(\HH )$ = $9.9 \times 10^3 \, f_{\rm D}^{1.4}$ (in the density 
range between 3$\times$10$^4$ \percc \ and 10$^6$ \percc), 
as found from a linear least square fit to the observed $f_{\rm D}$ data in 
Fig.~\ref{fd_dust} and the $n(\HH )$ data derived in Sect.~\ref{sdensity}
from the 1.2~mm continuum observations.
This allows us to find a 
simple relation for the deuterium fractionation as a function 
of $f_{\rm D}$ in ``L1544--like'' cores, if one neglects multiply deuterated
forms of \HTHP \ (see below):
\begin{eqnarray}
R_{\rm deut} & \equiv & \frac{[\NTDP ]}{[\NTHP]} \simeq 
	\frac{1/3 [\HTDP ]/[\HTHP ]}{1 + 2/3[\HTDP ]/[\HTHP ]} ,   
\label{erdeut}
\end{eqnarray}
and
\begin{eqnarray}
\frac{[\HTDP ]}{[\HTHP ]} & = & \frac{k_{\rm HD} [{\rm HD}]}
	{k_{\rm CO} [{\rm CO}] + k_{\rm \MOLN} [\MOLN ] 
	+ k_e [e] + k_{gr} [gr]},
\label{erd_general}
\end{eqnarray}
where $k_{\rm HD}$ is the rate coefficient of reaction ~\ref{eh2dp}
(see below), 
[HD]/[\MOLH ] $\equiv$ 2$\times$[D]/[H] = 3$\times$10$^{-5}$ (Oliveira et al.
\cite{ohk03}), 
$k_{\rm CO}$ = 3.3$\times$10$^{-9}$ $(T/10{\rm K})^{-0.5}$ cm$^3$ s$^{-1}$ 
is the rate 
coefficient of the reaction \HTDP \ + CO $\rightarrow$ $products$ ,
$k_{\MOLN}$ = 1.7$\times$10$^{-9}$ cm$^3$ s$^{-1}$ is the rate 
coefficient for the reaction \HTDP \ + \MOLN $\rightarrow$ $products$, 
[\MOLN ]/[\MOLH ] = 2$\times$10$^{-5}$ (Lee et al. \cite{lbh96}), 
$k_e$ = 5.5$\times$10$^{-7}$ $(T/10{\rm K})^{-0.65}$ cm$^3$ s$^{-1}$ is the dissociative 
recombination rate of \HTDP \ (see Caselli et al. \cite{cwt98}), and 
$k_{gr}$
is the rate of recombination of \HTDP \ on negatively charged dust grains, 
which, following Draine \& Sutin (\cite{ds87}), is given by:
\begin{eqnarray}
k_{gr} ({\rm cm^3 s^{-1}})& = & 1.6 \times 10^{-7} \frac{a_{\rm min}}{10^{-8} {\rm cm}}
	\left( \frac{T}{10 \, {\rm K}} \right)^{-0.5} \nonumber \\  
       & & 
      \times \left( 1 + 3.6 \, 10^{-4} \frac{T}{10\, {\rm K}}
	\frac{a_{\rm min}}{10^{-8} {\rm cm}} \right) .
\end{eqnarray}
Note that $k_{gr} \times [gr]$ is equivalent to $\alpha_{gr}$ in 
 Draine \& Sutin (\cite{ds87}), where $[gr]$ = $n_{\rm gr}/n(\MOLH )$ 
is the fractional abundance of dust grains and $n_{\rm gr}$ = 
$\int n_{\rm gr}(a) da$.
Introducing numerical values into eq.~\ref{erd_general}, we obtain
an expression for [\HTDP ]/[\HTHP ] which only depends on $f_{\rm D}$ 
and $a_{\rm min}$:
\begin{eqnarray}
\frac{[\HTDP ]}{[\HTHP ]} & = & 
	\frac{(k_{\rm HD }/1.5 \times 10^{-9} {\rm cm^3 s^{-1}})}
	{7/f_{\rm D} + 2.8 + 1.6/f_{\rm D}^{0.7} + 37.8 \, f(a_{\rm min})} , 
\label{erd}
\end{eqnarray}
where
\begin{eqnarray}
f(a_{\rm min}) & = & \left( \frac{k_{gr}}
	{1.6 \times 10^{-7}{\rm cm^3 s^{-1}}} \right) \left( 
	\frac{a_{\rm min}}
	{10^{-8} {\rm cm}} \right)^{-2.5} .
\end{eqnarray}
The last term in the denominator of eqn.~\ref{erd} 
can be neglected
as soon as $a_{\rm min}$ = 0.05 $\mu$m or larger, so that,  
for molecular ions, dissociative recombination 
becomes more important than recombination on grains. Hence, 
it can be neglected in dense clouds, where small grains are likely 
to be depleted on 
the surface of large grains or coagulated in larger fluffy structures 
(Ossenkopf \cite{o93}).  Depletion itself causes $a_{\rm min}$ $\approx$
0.01~$\mu$m (Walsmley et al.~\cite{wfp03}). 

The value of $k_{\rm HD }$ 
typically used in chemical codes is 1.5$\times 10^{-9}$ cm$^{3}$s$^{-1}$, 
which 
we call the ``standard rate'' (see e.g. Roberts et al. \cite{rhm03}).  However, Gerlich et al. (\cite{ghr02})
have recently measured a lower value for this rate (3.5$\times 10^{-10}$ 
cm$^{3}$s$^{-1}$, the ``GHR rate'') and in Fig.~\ref{rd_fd} results obtained
with the ``standard'' and ``GHR'' rate are shown. 
The observed data points lie between the two curves,
and the best fit (dotted curve in bottom figure) 
is obtained with $k_{\rm HD}$ = 7.5$\times$10$^{-10}$ cm$^{3}$ 
s$^{-1}$, which may suggest that a rate slightly (factor or $\sim$ 2) 
larger than the one recently 
measured is probably needed to explain observations.
However, one should note that in the case of L1544 ($R_{\rm deut}$ $\sim$ 
0.24 at $f_{\rm D}$ $\sim$ 10), even the ``standard rate'' fails 
to reproduce the large deuterium fractionation observed toward the dust peak.  
The only way to reach $R_{\rm deut}$ $\sim$ 0.24 with this analytical 
model (dash--dotted curve in Fig.~\ref{rd_fd}) is to allow a drop in the 
abundance of \MOLN \ in the central 
regions where $f_{\rm D}$ $\ge$ 10 and include in the chemical scheme
D$_2$H$^+$(\footnote{When  D$_2$H$^+$ is included in the model, 
equation \ref{erdeut} becomes:
\begin{eqnarray*}
R_{\rm deut} & \simeq 
	\frac{1/3 [\HTDP ]/[\HTHP ] + 2/3 ([\HTDP ]/[\HTHP ])^2}
	{1 + 2/3[\HTDP ]/[\HTHP ] + 1/3 ([\HTDP ]/[\HTHP ])^2} .
\end{eqnarray*}}), which becomes abundant in heavily (CO, \MOLN , and O) 
depleted regions (Roberts et al. \cite{rhm03}; Walmsley et al. \cite{wfp03}).  
Therefore, the difference in deuterium fractionation between L1521F and L1544 
is likely to be due
to a different evolutionary stage, with L1521F being less
evolved than L1544 (see also Aikawa et al. \cite{aoh03}).  
The inclusion of the reaction \HTDP \ + HD $\rightarrow$ D$_2$H$^+$ + \MOLH \
(Gerlich et al. \cite{ghr02}) in this model 
leads to an increase of the deuterium fractionation  by a factor of 2--3. 
However, in this simple chemical scheme we did not include the 
so--called ``back'' reactions due to ortho--\MOLH 
\ (see Gerlich et al. \cite{ghr02}), which have the effect of reducing the 
deuteration degree (see Walmsley et al. \cite{wfp03}); this analysis is beyond
the scope of the present paper.  

\subsubsection{A simple chemical model in a centrally concentrated core}
\label{sconcentrated}

 The analytic calculation outlined in the previous section has
 at least two major defects. One is the assumption of no abundance
variations within the dense core, which should be computed first 
in order to estimate molecular abundances as a function of radius and
then determine the column densities via integration along the line of 
sight. The second is our neglect of reactions
with species such as O which also act
to limit deuterium fractionation. Here, we present a small toy
model aimed at overcoming these problems, and already described in 
 Caselli et al. (\cite{cas02b}).
This model follows the (steady--state) chemistry of a spherically 
symmetric cloud with a density profile deduced from the 1.2mm dust continuum
emission (see Sect.~\ref{sdensity}), and constant temperature 
$T$ = 10 K. Neutral species in the model are \MOLH , CO, \MOLN , and 
atomic oxygen.  We follow depletion of these species onto dust grains and 
their desorption due to the cosmic ray impulsive heating of the dust, 
following the procedure by Hasegawa \& Herbst (\cite{hh93}).  
The abundance of molecular ions such as 
\HCOP , \NTHP \ and corresponding deuterated isotopomers are given by the 
steady state chemical equations using the istantaneous abundances of neutral
species.    Analogously, the electron 
fraction $x(e)$ can be computed in terms of global estimates for the molecular
and metallic ions and using the same istantaneous abundances of CO, \MOLN , 
and O in the gas phase.  The approach we have adopted is a simplified 
version of the reaction scheme of Umebayashi \& Nakano (\cite{un90}), 
which includes
molecular ion recombination on grain surfaces using rates from Draine 
\& Sutin (\cite{ds87})  
 (see Caselli et al.~\cite{cas02b} for details).
 
This model furnishes the abundances of gaseous species as a 
function of radius, and the corresponding column densities as 
a function of impact parameter are calculated taking into account 
the appropriate beam convolution to simulate the observations.
We used the new value of the dissociative recombination rate for 
\HTHP \ (4$\times$10$^{-7}$ cm$^{3}$ s$^{-1}$ 
at 10 K) determined by McCall et al. (\cite{mhs03}), assumed $k_{\rm HD}$ = 
3.5$\times$10$^{-10}$ cm$^{3}$ s$^{-1}$ (see previous section)
and varied other parameters
(essentially the binding energies on grain surfaces of \MOLN \ and O). 
The model has been
forced to reproduce within 10\% the observed \CSEO , \NTHP , and 
\NTDP \ column densities toward the dust peak position 
(5.5$\times 10^{14}$, 1.6$\times 10^{13}$, 
and 1.5$\times 10^{12}$ cm$^{-2}$, respectively), and to 
reproduce within a factor of 2 the observed column density profiles in 
the above molecules.  The best fit binding energies for \MOLN \ and O (800 K and 750 K, 
respectively) are quite close 
to the values deduced from theoretical calculations and laboratory measurements
(800 K and 750 K, for \MOLN \ and O, respectively; 
see discussion in Hasegawa et al. \cite{hhl92} 
and Bergin \& Langer~\cite{bl97}).

As illustrated in Fig.~\ref{abundances}, 
the fractional abundance of \NTHP \ decreases from about 
 2$\times$10$^{-10}$ at the edge of the cloud to about 
 1$\times$10$^{-10}$ at the 
center.  The increase of the \HTDP /\HTHP \ abundance ratio 
toward the center boosts the formation of \NTDP , which presents
an abundance increase of about one order of magnitude from the edge 
 (where $x(\NTDP )$ = 3$\times$10$^{-12}$) to the central 4000 AU 
($x(\NTDP )$ = 2$\times$10$^{-11}$) of the core.  
Therefore, in the case of L1521F, we do not need to 
invoke any ``molecular hole'' as in the case of L1544 (see Caselli et al.
\cite{cvc03}), although the present data do not allow us to distinguish 
between the models with different amounts of \MOLN \ freeze out 
(and consequent \NTHP \ depletion) within few thousands AU. 
On the other hand, recent observations of {\it ortho}-\HTDP \ toward this
object (Caselli, van der Tak, Ceccarelli et al., in preparation) 
strongly suggest that the molecular hole in L1521F must be smaller 
than in L1544, 
given that the line is about two times less intense than in L1544.  Indeed, 
we predict here an \HTDP \ abundance of 3$\times$10$^{-10}$ 
at radii less than 3000 AU, a factor of about 3 lower than in L1544
(see Caselli et al. \cite{cvc03}).  
 This is another indication that L1544 is likely to be more evolved,
and more 
centrally concentrated, than L1521F.  This seems to contradict the 
observational evidence that the central densities in the two cores 
are quite similar.  However, one should keep in mind that the  
central density values (and the density structure) in the two cores have been 
obtained assuming isothermal conditions.  Allowing the temperature to 
decrease toward the center, as predicted by models of dense clouds 
heated by the external radiation field (e.g. Galli et al. \cite{gal2002})
one finds that lower values of the central temperature
implies larger central densities (e.g. Evans et al. \cite{ers01}; 
Zucconi et al. \cite{zuc01}).  
One possible solution to the L1521F/L1544 puzzle is that 
L1544 is colder and more centrally concentrated than L1521F and 
hence is more depleted in the nucleus.  This is 
consistent with L1544 being more evolved and closer to the ``critical'' 
state than L1521F.  

We note that the chemical composition shown in 
Fig.~\ref{abundances} reproduces
the observed $R_{\rm deut}$--$f_{\rm D}$ (see the thick 
curve in Fig.~\ref{rd_fd})
and  $f_{\rm D}$--$N(\MOLH )$ (see the thick curve in Fig.~\ref{fd_cd})
relations, without any need to change the value of $k_{\rm HD}$ (see previous
section). This demonstrates the importance of taking into account core
structure and differential molecular freeze--out in chemical models.  
Moreover, note that the depletion factor  {\it within} the cloud, 
$F_{\rm D}$, is significantly larger 
(more than two orders of magnitude) than the integrated 
CO depletion factor $f_{\rm D}$, which is ``diluted'' along the line of sight 
(compare $F_{\rm D}$ in Fig.~\ref{abundances} with $f_{\rm D}$ in 
Fig.~\ref{fd_cd}).  Finally, the central value of the electron fraction is 
$\sim$5$\times$10$^{-9}$, implying an ambipolar diffusion time scale 
(see e.g. Shu et al.~\cite{sal87}) of $\sim$
2.5$\times$10$^{13} x(e)$ = 1.2$\times$10$^{5}$ yr, only slightly (factor of 
$\sim$3) larger than the free--fall time scale, 
once again suggesting that the core is close to dynamical collapse 
(although not as 
close as L1544).  As in L1544, the major ion in the core 
is H$_3$O$^+$ , which is due to the presence of a significant fraction of 
gaseous atomic oxygen in the model (see also Aikawa et al. \cite{aoi01} for 
similar results).  We should however point out that the more recent 
models of Aikawa et al. (\cite{aoh03}), where surface chemistry is included, 
predict a much lower abundance of O in the gas phase, given that 
in this model, an O--atom sticking to a grain  
is converted to water, which remains on the surface (assuming desorption 
due to the cosmic--ray impulsive heating of dust grains; 
see Hasegawa \& Herbst \cite{hh93}). Low ionization potential 
elements, essentially S$^+$, Mg$^+$, Fe$^+$, Si$^+$, Na$^+$, generically
called ``metals'' (M$^+$ in 
Fig.~\ref{abundances}), are assumed to freeze out onto dust grains at the
same rate as CO molecules.  For this reason, they are negligible carriers 
of positive charges within the core, in our model.

\subsection{Kinematics}   
\label{skinematics}

In order
to facilitate the comparison between L1521F and L1544, we analyzed the
kinematical properties of L1521F using the same models as in Caselli et 
al. (\cite{cas02a}). 
In particular, starting from the velocity field predicted by the
ambipolar diffusion models of Ciolek \& Basu (\cite{cb2000}; 
hereafter CB), we have derived the linewidth
gradient and the line profile in a disk--like geometry and compare the
results with our observations.

From the analysis of \NTHP (1--0) data,   
we found that the line width, similarly to L1544, 
decreases with distance from the dust peak (see Fig.~\ref{dv_b}). As seen in 
Caselli et al. (\cite{cas02a}), this observational evidence is consistent with the 
predictions of the CB model.  Here, we applied the 
kinematic analysis of L1544 (Caselli et al.~\cite{cas02a}) to the case of L1521F, 
assuming that 
the cloud has a disk--like shape and is centrally concentrated, with 
the center coincident with the 1.2mm dust continuum map peak.
From the core axial ratio, and following equation (1) of CB, 
the angle between the vertical axis of the model and the plane of the sky 
is found to be 18\deg .  The disk is contracting via ambipolar diffusion
and the resultant velocity field is used as input in a model which 
computes 
synthetic profiles of optically thin lines for all lines of sight across
the model disk (for details, see Caselli et al.~\cite{cas02a}).  As in the case of 
L1544, we also assumed that the density profile and the radial velocity 
field is given by the CB model at time $t$ = 2.66 Myr ($\equiv t_3$
in CB), which best reproduces the continuum observations of both cores. 
Line broadening is both due to thermal motions ($\sigma_{\rm th}$ =
0.05 \kms \ at 10~K) and microturbulence described by a turbulent velocity 
$\sigma_{\rm tu}$ independent of position.

Fig.~\ref{min+maj} shows the comparison between observed (hystogram) and 
synthetic (curves) profiles of the \NTHP (1--0) isolated hyperfine component 
($F_1F$ = 01$\rightarrow$12) along the minor and major axes of the L1521F core.
The (optically thin) weakest component ($F_1F$ = 10$\rightarrow$11) shows very 
similar profiles; so that we decided to present the higher sensitivity 
observations of the isolated component. We also considered two different 
conditions in the model: (1) the \NTHP \ abundance is constant throughout the 
core, so that the \NTHP \ column density simply follows the dust, and 
(2) there is a ``hole'' (2000 AU in size) in the
\NTHP \ distribution.  The difference between the two 
synthetic profiles is not significant (only a 4\% increase of the linewidth
toward the center, in the presence of the molecular hole, see 
Fig.~\ref{dv_b}), after the convolution with a 
gaussian with $\sigma$ = $\sqrt{\sigma_{\rm th} + \sigma_{\rm tu}}$ = 0.11 
\kms , needed to match the intrinsic 
linewidth toward the center\footnote{The linewidth toward the center and 
the observed $\Delta v$--$b$ trend can also be matched assuming 
a central infall speed 1.5 times larger than the one calculated 
by CB in their model $t_3$ and no extra broadening (e.g. turbulence)
needs to be invoked.  However, this is not consistent with the CB model
at time $t_3$ and we do not further discuss this possibility. On the 
other hand, recent \HTDP \ observations toward L1544 (Caselli et 
al.~\cite{cvc03}) also imply larger central infall speed than 
predicted by CB--$t_3$ 
 (van der Tak et al.~\cite{vcw03}).}, and thus only one profile 
is shown in Fig.~\ref{min+maj}. Also shown in the Fig.~\ref{min+maj}
are model profiles for the limiting case of no turbulent or thermal 
broadening (filled histograms).
The thing to note is that the agreement with the data is mixed, in the 
sense that the predicted velocity gradient along the minor axis is observed
but it is restricted to the 
south--west half of the axis. Along the major 
axis there is no clear gradient in the north--west half of the axis, 
as expected in absence of disk 
rotation, but a (0.06 \kms ) blue shift of the line is present toward
south--east. 

The synthetic profiles become narrower as we move away from the central 
40\arcsec \ of the disk--like cloud, given that the inward velocity in the 
adopted $t_3$ model reaches its maximum (0.12 \kms ) at radius 0.025 pc 
(or 37\arcsec ) before rapidly dropping to 0 at the cloud edge (see Fig. 2 of
CB).  We have analysed this line width variation to check its consistency with 
the observed trend shown in Fig.~\ref{dv_b}.  The solid curve in 
Fig.~\ref{dv_b}
is the CB prediction in the case of no central ``hole'', whereas the 
dashed curve illustrates the effects on the line width of a 
molecular ``hole'' in the central 2000 AU (i.e. a region where all 
heavy elements have condensed onto dust grains). 
The line width of model lines has been calculated as the
second moment of the velocity profile and it has been ``normalized'' to 
the value of the line width observed in the central position by 
convolving the purely--kinematic profiles with a Gaussian with $\sigma$ = 0.11
km s$^{-1}$, which can be interpreted {as the combination of thermal 
broadening plus} a constant turbulent field across 
the cloud. The comparison between the curves and the big dots 
(the binned data, see Sect.\ref{sdv-b}) suggests that our data are 
consistent with 
the CB model. In particular, the correspondence between the solid curve 
and the data indicates that the molecular hole is not present in L1521F, in 
agreement with the chemical analysis (see previous section).  On the other 
hand, the steeper $\Delta v$--$b$ relation found in L1544 (see Fig.~5 of 
Caselli et al.~\cite{cas02a}) suggests that the molecular hole is likely to be 
present in that source, again in agreement with the chemical analysis 
(see e.g. Caselli et al. \cite{cvc03}). 


As shown in Sect.~\ref{svel_field}, the kinematics of L1521F is also 
characterized by complex motions which may be the result of turbulence 
(see e.g. Burkert \& Bodenheimer \cite{bb00}) or 
accretion of material onto the core or a combination of both.  
Local gradients have been determined in order to quantify these motions, 
and we found that the magnitude of local gradient vectors tends to 
increase toward the core center.  
This suggests that turbulence (expected to dissipate more rapidly 
in denser environments), is probably not the driving 
source of the observed velocity field.  Moreover, unresolved substructure may 
further complicate the velocity field, as suggested by the \NTHP \ clumps 
observed by Shinnaga et al. (\cite{sol03}) (see Fig.~\ref{fhgrad}):  
with the exception of their 
clump ``N4'', the kinematics of the other clumps (N1--N3) is 
consistent with the velocity field inferred from our local gradient 
maps (see also Sect.~\ref{integrated})
and the (marginal, see Fig.~\ref{n2h+32}) 
evidence for two velocity components in \NTHP (3--2),
resembling clumps N1 and N2). 

It is interesting to compare our data with predictions from the
non--magnetic turbulent models of Ballesteros--Paredes et al. 
(\cite{bkv03}),
who derived velocity profiles for dense cloud cores. As an 
example, Fig.~\ref{fcuts} shows the velocity cuts observed across 
L1521F.  Within the half maximum contour of \NTHP (1--0), the 
{\it largest} velocity variation observed is $\la$ 0.04 \kms \
on a scale of 0.027 pc, corresponding
to a velocity gradient of $\la$ 1.5 km/s/pc.  On the other hand, 
Ballesteros--Paredes et al. 
(\cite{bkv03}) found that the {\it smallest} velocity variation 
for their clump 13
at time $t1$ (see their Fig.~9) is $\ga$ 0.3 \kms \ within 0.15 pc
or $\ga$ 2 km/s/pc.  Thus, current supersonic turbulent models predict 
velocity gradients which are somewhat too large. 
We also note that the reversal in the velocity gradient direction
observed in L1521F (see bottom panel of Fig.~\ref{fcuts}) 
is not present in the model examples shown in 
Ballesteros--Paredes et al. (\cite{bkv03}), but it is predicted 
by the turbulent models of Burkert \& Bodenheimer (\cite{bb00}).
 
In the two starless cores L1498 and L1517B, Tafalla et al. (\cite{tmc03})
found a good correlation between the distribution of CS and the
distribution of the high velocity \NTHP(1--0). 
The authors argue that the high velocity wing of the \NTHP(1--0) lines 
comes from a gas shell that is being accreted by the starless core; 
therefore it has not experienced the high density in the core nucleus 
and hence its depletion factor is low.
We repeated the same experiment in L1521F producing channel maps 
of the core in \NTHP(1--0) but we did not find any strong deviation from 
the total intensity distribution.
However, we did find that the \NTHP(1--0) velocity pattern across the core 
is similar to the \CEIO \ distribution.
In fact, as shown in Fig.~\ref{vel_co}, \NTHP(1--0) presents 
red--shifted velocities where CO is bright.
The difference in magnitude of this effect here and in L1517B and 
L1498 could be due to a brighter core that saturates the emission 
from the high velocity wing.  

A final thing to note is that 
Fig.~\ref{vel_co}, together with Fig.~\ref{fhgrad} (top figure), 
suggests the presence of a coherent velocity field resembling 
low-frequency spatial oscillations of the outer
cloud layers around some equilibrium dynamical state
as recently proposed by Lada et al. (\cite{lba03}) 
in the case of the starless core B68. However, L1521F is more massive
than B68 and is approaching the critical state for the onset of collapse, 
so that a situation of near--equilibrium for L1521F may be due to a balance 
between gravity and a combination of magnetic and thermal forces.  If 
this is the case, the normal modes for L1521F would be more complicated
than in the purely thermal case of B68, since the spherical symmetry
that is a fair approximation for a thermal--pressure supported 
cloud such as B68 will no longer be valid for a magnetically supported
object (G. E. Ciolek, {\it priv. comm.}). In fact, we observe L1521F 
not to be spherically symmetric (see Tab.~\ref{gauss2d}).

\subsection{Other possible interpretation}

Finally, in this section we discuss some more speculative 
interpretations of the evolutionary state of L1521F.  Firstly,
it seems that L1521F 
is likely to be in an earlier stage of evolution than L1544,
which is probably colder ($\sim$ 7 K in the central $\sim$ 2000 AU, see 
e.g. Zucconi et al. \cite{zuc01}) and more centrally concentrated (central 
densities $\sim$ 10$^7$ \percc ; e.g. Evans et al. \cite{ers01}) than L1521F, 
and hence more depleted in the nucleus. For example, in the CB model, 
this implies that whereas the density profile of L1521F is consistent
with the cloud model at time $t_3$=2.66 Myr, L1544 is closer to 
$t_4$=2.68 Myr, and thus the estimated age difference is roughly 20,000 yr.
However, these age estimates should be taken with caution given that
the two cores may have had a different formation and contraction history
and that, in general, cores, like stars, can differ in 
properties such as size, mass, and temperature regardless of their relative
age.

An alternative view can be based on the clumpy structure observed 
in L1521F (Shinnaga et al.~\cite{sol03}), but not in L1544 (e.g. Williams et 
al.~\cite{Wil99}), using BIMA. If the more complex kinematics in L1521F is 
due to the unresolved clumpy substructure, one may speculate that 
L1521F is close to the formation of a group of low mass 
protostars, whereas L1544 is likely to form one or two stellar objects.  
Thus, the two cores may show chemical and physical properties 
characteristic of the initial conditions of different modes of star 
formation in low mass cores.  One should also 
note that L1521F resides in the middle of the main filament of the 
Taurus Cloud, whereas L1544 is at the edges of the complex and hence 
the different environments of the two cores may be
responsible for the different kinematics and chemical properties 
of apparently similar (in the dust continuum and \CEIO \ emission) dense cores.

\section{Conclusions}

We have analysed physical and chemical properties of L1521F, a starless
core in the Taurus Molecular Cloud, with characteristics similar 
to the pre--stellar core L1544.  The main similarities and 
differences between the two cores are listed below:

1.  the dust emission distributions 
are similar, implying 
a fairly closely matched density structure, with central densities 
of $n_0$ $\sim$ 10$^6$ \percc , the radius of the ``flat'' region 
$r_0$ = 20\arcsec , and similar asymptotic power index $\alpha$ 
(see Sect.~\ref{sdensity} and Tafalla et al.~\cite{taf02}).  
In particular, the aspect ratio is quite similar: 1.6 and 1.8 in L1521F
and L1544, respectively.

2. The line width decreases with distance from the cloud 
center ($\sim$ 0.3 \kms ) to 
  80\arcsec \ away ($\sim$ 0.25 \kms ; see Fig.~\ref{dv_b}), in analogy 
with L1544, and consistent with the predictions of ambipolar diffusion
models, although any
gravity--driven contraction in 2-3D is expected to get localized line
broadening. The particular model which best fits the data will be
investigated in the future. 

3. The amount of CO freeze--out (integrated CO depletion factor 
$f_{\rm D}$ = 15) is also comparable to L1544, as is the column density of 
\NTHP \ toward the dust peak ($\simeq$ 1.5$\times$10$^{12}$ cm$^{-2}$). 

4. The deuterium fractionation toward the L1521F dust peak ($R_{\rm deut}$ 
$\sim$ 0.1) is however 
a factor $\sim$ 2 smaller than in L1544, due to the (factor of 2) 
smaller column density of \NTDP \ toward L1521F.  This can be understood
if L1521F is less chemically evolved than L1544, with a smaller
($r < 2000$ AU) central molecular hole. 

5. Unlike in L1544, the velocity field in L1521F maintains a 
complex structure even at the small scales traced by \NTDP \ and \NTHP (3--2) 
(see Figs.~\ref{fhgrad},\ref{fdgrad}).  This may be due to 
the presence of clumps in the central few thousand AU (as deduced 
by the interferometric observations of Shinnaga et al.~\cite{sol03}), but 
could also be caused by 
 normal mode pulsations, as in the case of B68 studied by
Lada et al. (\cite{lba03}). The ambipolar diffusion model with infall of 
Ciolek \& Basu (\cite{cb2000}) has problems in reproducing the whole velocity 
field observed across the core. This may be due to the fact that
part of the observed bulk motions result from residual core contraction,
as suggested by Tafalla et al.~(\cite{tmc03}) in their study of L1517B 
and L1498, thus preventing a clear determination of the
velocity field within the core nucleus.

6. The line profiles in L1521F show asymmetric structure, 
although the two peaks are not well separated as in L1544. This 
is consistent with smaller (factor of ~1.5) infall velocities in the 
central few thousand AU of L1521F. 

In any case,
the large central density ($\sim$ 10$^6$ \percc ) and the evidence of 
central infall (broader line widths toward the center and central infall 
asymmetry in \NTHP (1--0)) indicate that L1521F
is another starless core on the verge of star formation, or a
pre--stellar core.  
Although a study of a more complete sample is needed, assuming that L1544 and 
L1521F are the only two cores in Taurus close to dynamical collapse, and given 
that the total number of starless cores in Taurus is 44 (Onishi et al. 
\cite{omk02}),
we can argue that this process of contraction towards the ``critical'' 
stage, or the ``L1544--phase'', may last about five percent of the core 
lifetime.

A more detailed analysis of \NTHP \ line profiles will be presented in 
a future paper, where the observed chemical abundances will be introduced
in a non spherically symmetric
Monte Carlo radiative transfer code. 
This is needed to understand the
nature of the \NTHP (1--0) line asymmetry, which may be caused by
self--absorption plus infall, or to infall plus a molecular hole, 
or to the relative motion of different clumps, or to a 
mixture of the above possibilities.   

\begin{acknowledgements}
The authors are grateful to the staff of the IRAM 30m antenna, for their
support during observations, and to Richard Crutcher and Daniele Galli
for discussion.  We also thank the referee, Glenn Ciolek, for 
clarifying
several statements in the paper and Floris van der Tak for a critical
reading of the submitted manuscript.
PC and CMW aknowledge support from the MIUR project ``Dust and Molecules 
in Astrophysical Environments''. CWL acknowledges support
from the Basic Research Program (grant KOSEF R01-2003-000-10513-0) of
the Korea Science and Engineering Foundation.
\end{acknowledgements}


\begin{figure}[!htbp]
 \caption{Dust emission from L1521F.  Levels are 30, 55 and 80 mJy/beam. Reference position  is 
 RA: 04:28:39.8 DEC: 26:51:35 (J2000). 
 The dashed ellipse best fits the core structure with a 2D gaussian. 
The dotted ellipse is the result of a 2D--gaussian fit to the whole map, including 
the more extended emission. The black rectangle shows the 
 area mapped in line observations (see Fig.~\ref{fig1b}).}
 \label{fig1a}
\end{figure}

\begin{figure}[!htbp]
 \caption{L1521F integrated intensity maps in the observed  molecular transitions. Contour levels are 
 45, 70, 95\% of the relative peak in each map (whose values are: 5.9,  0.74,  1.1,   0.33,   2.5 and  2.3 \kks in  \NTHP(1-0),
 \NTHP(3-2),\NTDP(2-1),\NTDP(3-2),\CEIO(1-0) and \CEIO(2-1) respectively).
 The \NTHP(3-2) data were smoothed to 26\sec resolution to increase S/N and help the comparison with 
 \NTHP(1-0); for the same reasons \NTDP(3-2)
 was smoothed to 16\sec \ (the HPBW of the 30m antenna at the frequency of 
the the \NTDP (2--1) line). We overlaid on each map the ellipse which best fits the dust emission structure
 (in dashed line) and the 80 mJy/beam contour (dotted line).  }
 \label{fig1b}
\end{figure}

\begin{figure}[htbp]
\caption{Spectra of \NTHP (1-0),  
 \NTHP (3-2),  \NTDP (2-1), \NTDP (3-2), \CEIO (1-0), and \CEIO (2-1)
towards the dust emission peak in L1521F. A fit of the hyperfine
pattern (or in the case of \CEIO \ a gauss fit) of the lines assuming
identical excitation temperatures for all satellites is also shown
(dashed) for comparison.  }
 \label{spectra}
\end{figure}

\begin{figure}[htbp]
\caption{Enlargement of the \NTHP (3--2) spectrum of Fig.~\ref{spectra}
 toward offset (-10,10), 
which shows tentative evidence for a double--peaked feature.  The top 
panel show the line profile (grey histogram) and the hfs fit (black curve)
assuming two velocity components; the bottom panel show the hfs fit 
assuming one velocity component. The velocities of the two components 
concide with the LSR velocities of clumps N1 and N2 observed by
Shinnaga et al. (\cite{sol03}) in their interferometric observations, 
see Fig.~\ref{fhgrad} for reference.}
 \label{n2h+32}
\end{figure}

\begin{figure}[htbp]
 \caption{Circularly averaged dust emission 
fitted with the expression
 n$_0$/(1+(r/r$_0$)$^{\alpha}$) convolved with 
   a 10.5\sec gaussian. The parameters that best fit the L1521F data
   are $n_{0}= 10^6$ cm$^{-3}$,  $r_0 = 20$\sec and $\alpha = 2$. 
Dots with error bars are binned data and the dashed curve shows the 
observational beam.}
 \label{dprof}
\end{figure}

\begin{figure}[htbp]
 \caption{CO depletion factor (full contours) 
against dust emission smoothed to a 
22\sec beam (grey scale).  $f_{\rm D}$ contours range between 6 and 15, in 
steps of 3.  The maximum value of $f_{\rm D}$ is 18 and is located at offset
(0,9). Note the good correspondence between the distribution of \NTDP (2--1)
(dashed contours) and the CO depletion map.}
 \label{fd_dust}
\end{figure}

\begin{figure}[htbp]
\vspace*{1cm}
\caption{Integrated CO depletion factor $f_{\rm D}$ against \MOLH \ 
column density as derived from the 
data in Fig.~\ref{fd_dust}.  The dashed curve is the  
linear least fit to the data, whereas 
the solid curve is the $f_{\rm D}$
vs. $N({\rm H_2})$ relation found in the simple model described in 
Sect.~\ref{sconcentrated}.  \label{fd_cd}}
\end{figure}

\begin{figure}[htbp]
 \caption{[\NTDP]/[\NTHP] in the central region as estimated from CTEX 
calculations (see Sect.~\ref{sdeut}). 
Levels are 0.066, 0.078, and 0.091.  The dashed
contour is the small ellipse of Fig.~\ref{fig1b}.}
 \label{Dfrac_map}
\end{figure}

\begin{figure}[htbp]
\vspace*{1cm}
\caption{\NTHP (1-0) intrinsic line--width vs. impact parameter. The 
big dots are averages of the data points within 15\arcsec \ bins. The solid
curve is the $\Delta v - b$ relation derived from the velocity field
predicted by the $t_3$ model of  
Ciolek \& Basu (\cite{cb2000}) (see Caselli et al.~\cite{cas02a}),  
where a constant turbulent field across the cloud has 
been included, and the dashed curve is from the same model but 
accounting for the presence of 
a central molecular ``hole'' of 2000 AU in size (see text). 
\label{dv_b}}
\end{figure}

\begin{figure}[htbp]
\vspace*{1cm}
\caption{Isolated component profile along the P.A. = 45\deg axis passing  
through (-10,0) (left panel) and -45\deg axis passing through (-10,0) 
(right panel). This figure
shows how the linewidth increases going from the edge of the core to the
center. Fits are gaussian fits to this component.
\label{dv_phil}}
\end{figure}

\begin{figure}[htbp]
\caption{Velocity gradient vectors in L1521F derived from the two
\NTHP \ maps.  The integrated intensity of \NTHP (1--0) (top) and 
\NTHP (3--2) (bottom) is shown as grey scale. The thick contour 
marks the integrated intensity half--maximum contour.
The arrows across the map are ``local'' velocity gradients
obtained in a 3$\times$3--point grid of positions centered on the 
corresponding grid, indicating the magnitude and the direction. 
The arrow in the bottom right of the figure is the 
``total'' velocity gradient derived from the whole set of available
positions with $A/\sigma_A$ $>$ 10 ($A$ being the integrated 
intensity). Note the large variation in magnitude and direction
of the local gradients. The white circles in the top figure locate 
the positions of the four \NTHP \ clumps detected by Shinnaga et 
al.~(\cite{sol03}).}
\label{fhgrad}
\end{figure}

\begin{figure}[htbp]
\caption{Same as Fig.~\ref{fhgrad} for \NTDP \ data.  A similar 
pattern is present.}
\label{fdgrad} 
\end{figure}

\begin{figure}[htbp]
\vspace*{0.5cm}
\caption{Deuterium fractionation 
($R_{\rm deut} \equiv N(\NTDP )/N(\NTDP )$) as a 
function of integrated CO depletion factor.  Filled squares are 
data from this paper and the filled circle with error bar is our result
for the dust peak of L1544 (Caselli et al.~\cite{cas02b}).
Thin full curves are predictions from 
a simple chemical model using different rate coefficients for the 
proton--deuteron exchange reaction (\ref{eh2dp}):
the "standard" rate of 1.5$\times$10$^{-9}$ cm$^{3}$ s$^{-1}$
and the "GHR" rate of 3.5$\times$10$^{-10}$  cm$^{3}$ s$^{-1}$,
recently determined by Gerlich et al. (\cite{ghr02}).  Dashed curves refer
to the same models but with larger value of $a_{\rm min}$, the 
minimum radius of dust grains in the adopted MRN grain--size
distribution (see Sect.~\ref{ssimple}). 
The dotted curve is the best fit to the data, 
found if $k_{\rm HD}$ = 7.5$\times$10$^{-10}$ cm$^3$ s$^{-1}$. 
The thick curve is the result of 
a more comprehensive chemical model which takes into account the 
density structure of the core (see Sect.~\ref{sconcentrated}). 
Typical 1--$\sigma$ error bars are shown. To reproduce the $R_{\rm deut}$ value
observed in L1544, \MOLN \ freeze--out and multiply deuterated forms 
of \HTHP \ have to be included in the model (dash--dotted curve). \label{rd_fd}}
\end{figure}

\begin{figure}[htbp]
\caption{Fractional abundances, $n(i)/n(\MOLH )$,  as a 
function of radius in L1521F.  {The symbol M$^+$ refers to metals (see 
text).} The adopted density profile is 
the one described in Sect.~\ref{sdensity}.  This model 
 reproduces the observed column densities of CO, \NTHP , and 
\NTDP . The \NTHP \ abundance decreases towards the center by a 
factor of about 2, whereas \NTDP \ increases by a factor of $\sim$7 
from core edge to core center. 
Note that the \HTDP \ abundance is predicted to be $\sim$3$\times$10$^{-10}$
at the core center, a factor of 3 smaller than observed in L1544. 
In analogy with L1544, 
the \HCOP \ and \DCOP \ abundance profiles present a steep drop
within about 5000 AU.  The right axis refers to the CO depletion 
factor (see the curve labelled ``$F_{\rm D}$''). The set of 
parameters used to obtain this model are: $E_{\rm D}$(CO) = 1210 K,
$E_{\rm D}$(\MOLN ) = 800 K, $E_{\rm D}$(O) = 750 K, $\zeta$ = 
1.3$\times$10$^{-17}$ s$^{-1}$, $a_{\rm min}$ = 5$\times$10$^{-6}$ cm,
and $x(\MOLN )$ = 2$\times$10$^{-5}$,
where $E_{\rm D}(i)$ is the binding energy of species $i$ and 
$\zeta$ is the cosmic--ray ionization rate (see Caselli et al.~\cite{cas02b} 
for details of the model). 
\label{abundances}}
\end{figure}

\begin{figure}[htbp]
\caption{Observed line profiles of the $F_1\,F$ = 0~1$\rightarrow$1~2 
hyperfine component of \NTHP (1--0) (empty histograms) along the minor 
(left panels) and major (right
panels) axis of L1521F compared with predictions based on the 
Ciolek \& Basu~\cite{cb2000} velocity field
at time $t=t_3$ (full curves), convolved with a $\sigma$ = 0.11
\kms \ gaussian to reproduce the observed intrinsic linewidth of the central 
spectrum (see Fig.~\ref{dv_b}).  The filled histograms are 
model line profiles computed assuming no turbulent or thermal 
broadening. The core center is defined by the dust peak, located at 
offset (-8,+2).
\label{min+maj}}
\end{figure}

\begin{figure}[htbp]
\caption{Intensity (thin lines) and velocity (thick lines) 
of \NTHP(1-0) along four
different cuts; in the top panel continuous lines refer to the minor axis 
(PA 115\deg) while the dashed line refers to a cut at PA 70\deg;
 in the bottom panel,
the  continuous lines show the behaviour along the major axis 
(PA 25\deg) and the dashed
lines represent the other bisector (PA 160\deg).}
 \label{fcuts}
\end{figure}

\begin{figure}[htbp]
\caption{\CEIO (1--0) integrated intensity map (greyscale; wedge units
are K \kms ) overlaid
by the \NTHP (1--0) velocity map (contours). \NTHP \ velocity levels 
are from 6.37\kms ~to 6.55\kms spaced ~by 0.02\kms; higher velocity 
contours are represented as solid lines.  \label{vel_co}}
\end{figure}

\end{document}